\documentclass[11pt,a4paper]{article}
\usepackage{jstyle, enumitem}
\usepackage{amsmath,tabu}
\usepackage{hyperref}
\usepackage{slashed,framed}
\usepackage{empheq}
\usepackage{multirow}
\usetikzlibrary{calc,decorations.markings}

\DeclareMathOperator{\cosech}{cosech}
\DeclareMathOperator{\cosec}{cosec}
\def\l{\left}
\def\r{\right}
\def\s{\bar{s}}

\author[a]{Shailesh Lal}
\author[b]{\quad Prithvi Narayan}

\affiliation[a]{LPTHE -- UMR 7589, UPMC Paris 06, Sorbonne Universit{\'e}s,  Paris 75005, France}
\affiliation[b]{International Centre for Theoretical Sciences, Hesaraghatta,\\ 
Bengaluru North, 560 089, India}

\emailAdd{shailesh.hri@gmail.com}
\emailAdd{prithvi.narayan@gmail.com}
\begin{document}

\title{On Exponentially Suppressed Corrections to BMPV Black Hole Entropy}

\abstract{The microscopic formula for the degeneracy of BMPV black hole microstates contains a series of exponentially suppressed corrections to the leading Bekenstein Hawking expression. We identify saddle points of the quantum entropy function for the BMPV black hole which are natural counterparts to these corrections and discuss the matching of leading and next-to-leading terms from the microscopic and macroscopic sides in a limit where the black hole charges are large.
}

%\preprint{XX/YY}

\maketitle
%\flushleft
\section{Introduction}
The computation of the quantum entropy of black holes is a particularly challenging problem in quantum gravity. For BPS black holes in string compactifications that preserve sufficient supersymmetry, string theory provides an explicit statistical explanation for the origin of the entropy.
On taking the semi-classical limit by scaling the charges carried by the black hole one may recover the famous Bekenstein-Hawking area law, or its generalization, the Wald formula \cite{Wald:1993nt}. These computations date back to the original work of Strominger and Vafa \cite{Strominger:1996sh}, and have been very explicitly carried out in \cite{Dijkgraaf:1996it,Maldacena:1999bp,LopesCardoso:2004xf,Shih:2005uc,Jatkar:2005bh, Dabholkar:2006xa,David:2006ji,David:2006ru,David:2006yn,David:2006ud,Sen:2008ta,Sen:2009gy} subsequently, and provide important evidence for the viability of string theory as a theory of quantum gravity. We refer the reader to the reviews \cite{Sen:2007qy,Mandal:2010cj,Dabholkar:2012zz} for a more detailed account of these developments.

Interestingly, since the near horizon geometry of an extremal black hole always contains an AdS$_2$ factor \cite{Kunduri:2007vf,Figueras:2008qh}, one may use the AdS$_2$/CFT$_1$ correspondence to compute the the quantum degeneracy $d_{hor}$ associated with the horizon of an extremal black hole carrying charges $\vec{q}\equiv q_i$. This proposal, known as the quantum entropy function, states that \cite{Sen:2008vm,Sen:2009vz}
\begin{equation}\label{qef}
d_{hor}\l(\vec{q}\r)\equiv\left\langle\exp\l[i\oint q_i d\theta\mathcal{A}_\theta^i\r] \right\rangle_{\text{AdS}_2}^{finite}.
\end{equation}
Here $\mathcal{A}_\theta^i$ is the component of the $i^{\text{th}}$ gauge field along the boundary of the $\text{AdS}_2$, where $i$ runs over the set of all gauge fields in AdS$_2$, including those obtained by Kaluza Klein reduction. The path integral is carried out over all fields that asymptote to the black hole near horizon geometry. The superscript `\textit{finite}' indicates that the volume divergence in this path integral is 
regulated in the standard manner of the AdS/CFT correspondence.\footnote{
In particular, if we parametrize AdS$_2$ in global coordinates
\begin{equation}
ds^2 = a^2\left(d\eta^2 +\sinh^2\eta \,d\theta^2\right),\quad \eta\in\left[0,\infty\right),\quad \theta \in\left[0,2\pi\right),
\end{equation}
then we may regularize the AdS volume divergence by placing a cutoff $\eta_0$ on the AdS$_2$ radial coordinate $\eta$. Then in the limit where $\eta_0$ is large, we have
\begin{equation}
\ln d_{hor}\l(\vec{q}\r) =\mathcal{O}\left(e^{\eta_{0}}\right)+ \mathcal{O}\left(1\right) +\mathcal{O}\left(e^{-\eta_0}\right).
\end{equation}
Now the $\mathcal{O}\left(e^{\eta_{0}}\right)$ term is proportional to the length of the regularized AdS$_2$ boundary, and may be removed by the addition of local counterterms which have support only on the boundary of AdS$_2$. The second term, which is $\mathcal{O}\left(1\right)$, cannot be removed in this manner and should be regarded as physical. This leads us to the following renormalization prescription \cite{Sen:2008vm,Sen:2009vz}: First, we regulate the AdS$_2$ radial coordinate as above, and then compute the free energy associated to \eqref{qef} in the large $\eta_0$ limit. We then simply discard the term which diverges as $\mathcal{O}\left(e^{\eta_0}\right)$, then take the limit of $\eta_0$ going to infinity, which retains only the $\mathcal{O}\left(1\right)$ term.}

This proposal has already led to interesting insights into the quantum properties of four dimensional half-BPS black holes in $\mathcal{N}=2$ supergravities \cite{Sen:2011ba,Gupta:2014hxa} that were previously unavailable from microscopic analyses. Further, methods of supersymmetric localization have been brought to bear on this path integral with results that are promising for further investigation \cite{Dabholkar:2010uh,Dabholkar:2011ec,Gupta:2012cy,Dabholkar:2014ema,Murthy:2015yfa,Gupta:2015gga,Murthy:2015zzy,Gomes:2013cca}. 

In this paper we shall evaluate the quantum entropy function in a saddle point approximation which is valid in a particular scaling limit of the black hole charges.
In the context of four dimensional black holes, it is well known that the large-charge expansion of the microscopic degeneracy contains extra terms which are exponentially suppressed with respect to the leading Bekenstein-Hawking contribution \cite{Maldacena:1999bp,Banerjee:2008ky,Banerjee:2008pu,Dabholkar:2008zy}. 
Further, it was proposed that these terms may be interpreted in the QEF as arising from a class of saddle points of the quantum entropy function obtained by taking orbifolds of the near horizon geometry \cite{Banerjee:2008ky,Sen:2009vz,Sen:2009gy,Banerjee:2009af}. Further, it was shown that the matching persists when quantum effects are included in the AdS$_2$ side \cite{Gupta:2013sva,Gupta:2014hxa}, generalizing on the computations performed about the dominant saddle point \cite{Banerjee:2010qc,Banerjee:2011jp}

Motivated by these developments, we shall attempt to construct saddle-points of the quantum entropy function for five dimensional extremal black holes by taking particular $\mathbb{Z}_s$ orbifolds of the near horizon geometry by applying the results of \cite{Banerjee:2009af}. These will again give rise to contributions to $d_{hor}$ exponentially suppressed with respect to that from the near horizon geometry. Next we shall study the microscopic formulae for the degeneracy of these black holes and demonstrate that corresponding terms are indeed found in the microscopic formula in the same scaling limit for black hole charges, and this matching persists to the next-to-leading order in the large charge expansion. 

The next-to-leading order term is known as the \textit{log term}. It receives contributions only from one-loop fluctuations of massless fields about the saddle point. Further, its value is sensitive to only two-derivative terms in the quadratic action \cite{Sen:2012dw}. Importantly, this makes clear that the log term is an important window into the quantum properties of black holes. It is sensitive only to infrared physics, and yet carries information about the underlying microscopic theory that the black hole is placed in.

A brief overview of this paper is as follows. We begin with a summary of the properties of the near horizon geometry of the BMPV black hole and its non-rotating counterpart in Section \ref{sec:bmpvnhg}. We next turn to a construction of the saddle points of the quantum entropy function for the BMPV and Strominger-Vafa black holes in Section \ref{sec:qefsaddle}. Section \ref{sec:logterms} describes the computation of next-to-leading order corrections, known as log terms, about these saddle points. This concludes our macroscopic analysis. We next turn to the microscopic side where we first explicate the large-charge expansion of the microscopic degeneracy computed in Type II string theory on $T^5$ in Section \ref{sec:microt5} and match the results obtained with the predictions of Section \ref{sec:logterms}. Further details of computations may be found in the Appendices.

\section{The BMPV Black Hole and Its Near Horizon Geometry}\label{sec:bmpvnhg}
In this section we review the near horizon geometry of the BMPV black hole \cite{Breckenridge:1996is} embedded into Type IIB string theory compactified on ${\cal M} \times S^1$,
where  $\cal M$ is either $T^4$ or $K3$.\footnote{The analysis also extends to orbifolds $\left({\cal M} \times S^1\right)/ \mathbb{Z}_N$. Examples of such orbifolds are  the CHL orbifolds 
\cite{Chaudhuri:1995bf,Chaudhuri:1995dj,Chaudhuri:1995fk,Aspinwall:1995fw} and the orbifolds of \cite{David:2006ru,David:2006ud}. For this reason, in Section \ref{sec:logterms} we shall work with a generic number $n_V$ of massless $U(1)$ gauge fields in the effective five-dimensional supergravity. For the cases of immediate interest, i.e. ${\cal M} \times S^1$ compactifications, $n_V=27$.}. 
These black holes carry $Q_5$ units of D5 brane charge along ${\cal M} \times S^1$, $Q_1$ units of D1 brane charge along $S_1$, $-n $ units of momentum along $S^1$
, ${J \over 2}$ units of $J_{3L}$ charges, and finally zero $J_{1L},J_{2L},J_{1R},J_{iR}$ charges. Here the $J_{iL}$ and $J_{iR}$ are the generators of the $SO(4) \simeq SU(2)_L \times SU(2)_R$ rotational symmetry that a five dimensional black hole is charged under.

\subsection{The Near Horizon Geometry of BMPV Black Hole}
In this section we will describe the near horizon geometry of the BMPV black hole as a solution of Type IIB string theory on $\cal M$. However, since the internal directions ${\cal M}$ play no role in our computations, they are suppressed below. Dimensional reduction on the $S^{1}$ labelled by $\chi$ results in the usual BMPV black hole, whose near horizon geometry is paramatrized by the coordinates $\rho,\tau,x_{4},\psi,\phi$. For more details on the full 10 dimensional field configuration, we refer the reader to \cite{Dabholkar:2006tb,Banerjee:2009uk,Sen:2012cj}. The Lorentzian near horizon geometry of the black hole is given by 
\begin{equation}\label{eq:BMPVOld}
\begin{split}
    ds^{2} & =  r_{o}\frac{d\rho^{2}}{\rho^{2}}-r_{o}\rho^{2}d\tau^{2}+(d\chi-A\rho d\tau)^{2}+r_{o}(dx_{4}+\cos\psi d\phi-B\rho d\tau)^{2}\\
 &   +r_{o}(d\psi^{2}+\sin\psi^{2}d\phi^{2})+\frac{\tilde J}{4r_{o}}(d\chi-A\rho d\tau)(dx_{4}+\cos\psi d\phi-B\rho d\tau),
\end{split}
\end{equation}
with the Ramond Ramond 3 form flux taking the value
\begin{equation}
     \label{flux}
 F = {r_o \over \lambda} \left\lbrace \epsilon_3 + (*_6) \epsilon_3 + {\tilde J \over 8 r_0^2 } d\chi \wedge    \left[ \sin\psi d\psi \wedge d\phi +   {d\rho \over \rho} \wedge \left( dx^4 + \cos \psi d\phi  \right) \right]  \right\rbrace,
\end{equation}
where 
\begin{eqnarray}\label{ABparam}
A & = & \sqrt{r_{o}}\left[1-\frac{\tilde J^{2}}{64r_{o}^{3}}\right]^{-\frac{1}{2}},\qquad\text{and}\qquad B=-\frac{\tilde J}{8r_{o}^{2}}A.
\end{eqnarray}
Here 
$\epsilon_3\equiv \sin \psi dx_4 \wedge d\psi \wedge d\phi$ is the volume form on three sphere and $(*)_6$ denotes the hodge dual in six dimensions $\tau,\rho,x_4,\psi,\phi,\chi$. Notice that when $\tilde  J\ne0$, the BMPV metric does not factor
into $AdS_{2}\times S^{3}$. In fact both $S^{3}$ (parametrized by
$x_{4},\psi,\phi$) and $S_{1}$ (parametrized by $\chi$) are nontrivially
fibred over the base $AdS_{2}$. The periodicity is given by 
%Eq (4.2) of \cite{Sen:2012cj}) 
\begin{equation}
(\psi,\phi,x_{4})=(2\pi-\psi,\phi+\pi,x_{4}+\pi)=(\psi,\phi+2\pi,x_{4}+2\pi)=(\psi,\phi,x_{4}+4\pi)
\end{equation}
and 
\begin{equation}
\chi \equiv \chi + 2 \pi R_5,
\end{equation}
coupled with some identifications in $\cal M$. The various parameters appearing in the above metric $r_0,R_5,\tilde J$ are related to the string theory parameters 
$Q_1,Q_5,n,N,J$, and the string coupling $\lambda$, as
\begin{equation}
r_o = {\lambda Q_5 \over 4},\quad  R_5 = \sqrt{\lambda n\over N Q_1},\quad \tilde J = {J Q_5 \lambda^{3 \over 2} \over 2 \sqrt{Q_1n/N}},\quad V_{\cal M} \sim  {Q_1 \over Q_5},
\end{equation}
where $V_{\cal M}$ denotes the size of the internal manifold $\cal M$. 
We will work in the scaling limit
\begin{equation}\label{scalinggravity}
Q_1,Q_5\sim\Lambda, \quad n\sim\Lambda,\quad J\sim\Lambda^{\frac{3}{2}},\qquad \Lambda\rightarrow\infty.
\end{equation}
In this limit, we see that the components of the metric purely along the directions $\eta$, $\theta$, $\psi$, $\psi$ and $x_4$ scale as $r_o\sim {\cal O}(\Lambda^2)$, which is the dominant scaling behaviour. This effectively takes us to a five-dimensional limit of the six-dimensional geometry \eqref{eq:BMPVOld}.
It is easy to compute the Bekenstein-Hawking entropy for the black hole by computing the area of the horizon. We get
\begin{equation}
S_{BH} = \pi \sqrt{{4 Q_1 Q_5 n \over N}  - J^2} 
\end{equation}
The geometry has following Killing vectors: 
\begin{equation}\label{bmpvkilling1}
\begin{split}
    J_{1} & =  \sin\phi\partial_{\psi}+\cot\psi\cos\phi\partial_{\phi}-\cosec\psi\cos\phi\partial_{x_{4}},\\
J_{2} & = \cos\phi\partial_{\psi}-\cot\psi\sin\phi\partial_{\phi}+\cosec\psi\sin\phi\partial_{x_{4}},\\
J_{3} & = \partial_{\phi}, \quad L_{-}  =  \partial_{\tau}, \quad L_{0}  = \tau\partial_{\tau}-\rho\partial_{\rho},\\
L_{+} & =  \frac{1}{2}(\frac{1}{\rho^{2}}+\tau^{2})\partial_{\tau}-\tau\rho\partial_{\rho}+\frac{A}{\rho}\partial_{\chi}+\frac{B}{\rho}\partial_{x_{4}},\\
\hat{J}_{3} & =  \partial_{x_{4}}, \quad u = \partial_{\chi}.
\end{split}
\end{equation}
Note that the $L_{i}$ generate $SL(2,R)$ obeying 
\begin{equation}
[L_{0},L_{\pm}]=\pm L_{\pm}\hspace{20mm}[L_{+},L_{-}]=-L_{0}
\end{equation}
and $J_{i}$ generate $SU(2)_{L}$ i.e $[J_{i},J_{j}]=\epsilon_{ijk}J_{k}$
with $\epsilon_{123}=1$ and both $u$, $\hat{J}_{3}$ generate $U(1)_{u}$, $U(1)_{x_{4}}$ respectively.
Hence the isometry of the BMPV directions is $SL(2,R)\times SU(2)_{L}\times U(1)_{x_4}$. 
It can also be shown that the the above geometry has four killing
spinors. This results in the enhancement of symmetry (of the BMPV
directions) to the supergroup $SU(1,1|2)\times U(1)_{x_{4}}$. 
Also for later use, it is more useful to consider the following complex combinations of the Killing vectors 
\begin{equation}
k^{(1)} = J_1 +i J_2,\quad k^{(2)} = J_1 -i J_2,\quad k^{(3)} = J_3,\quad k^{(4)} = \hat{J}_3.
\end{equation}
Explicitly, these are given by
\begin{equation}\label{bmpvkilling}
\begin{split}
k^{(1)} &= e^{-i \phi }\left(i\,\partial_{\psi}+\cot\psi\,\partial_{\phi}-\cosec\psi\,\partial_{x_4}\right),\\
k^{(2)} &= e^{i \phi }\left(-i\,\partial_{\psi}+\cot\psi\,\partial_{\phi}-\cosec\psi\,\partial_{x_4}\right),\\
k^{(3)} &= \partial_{\phi}, \quad k^{(4)} = \partial_{x_4}.
\end{split}
\end{equation}
\subsubsection{The Special Case of $J=0$}\label{nonrotatingnhg}
It is also easy to see that when $J=0$ (and hence $B=0$), the $S_{3}$
is no longer fibred over AdS$_{2}$. The geometry has extra
Killing vectors, in addition to the Killing vectors listed in \eqref{bmpvkilling1}, which are given by 
\begin{equation}\label{extrakilling}
\begin{split}
\hat{J}_{1} & =  -\sin x_{4}\partial_{\psi}+\frac{\cos x_{4}}{\sin\psi}\partial_{\phi}-\cos x_{4}\cot\psi\partial_{x_{4}}. \\
\hat{J}_{2} & =  \cos x_{4}\partial_{\psi}+\frac{\sin x_{4}}{\sin\psi}\partial_{\phi}-\sin x_{4}\cot\psi\partial_{x_{4}} 
\end{split}
\end{equation}
Note that the $\hat{J}_{1},\hat{J}_{2},\hat{J}_{3}$, generate another
$SU(2)_{R}$. $J_{i},\hat{J}_{i}$ together generate the $SO(4)=SU(2)_{L}\otimes SU(2)_{R}$
isometry of the the three sphere $S_{3}$ and the supergroup is now enhanced
to $SU(1,1|2)\times SU(2)_{R}$. It is more useful to consider the complex combinations of the Killing vectors 
\begin{equation}\label{newkilling}
\begin{split}
\tilde{k}_1 &= \tilde{J}_1 + i\tilde{J}_2 = e^{i x_4}\left(i\,\partial_{\psi}+\cosec\psi\,\partial_{\phi}-\cot\psi\,\partial_{x_4} \right),\\
\tilde{k}_2 &= \tilde{J}_1 - i\tilde{J}_2 = e^{-i x_4}\left(-i\,\partial_{\psi}+\cosec\psi\,\partial_{\phi}+\cot\psi\,\partial_{x_4} \right).
\end{split}
\end{equation}

\subsubsection{The algebra of Isometry group }

When the black hole is dimensionally reduced to $AdS_{2}$ directions,
the above isometry group gets infinitely extended to affine algebra
$\hat{su}(1,1|2)$. It will turn out later that the computation of
entropy will involve the global symmetry charges of certain generators
of the above affine algebra. We give below the relevant part of algebra. 

Let $G_{n}^{\alpha\beta}$(where $\alpha,\beta\in1,2,\ n\in\mathbb{Z}+\frac{1}{2}$)
be the fermionic generators of the isometry algebra. As we will show
later, the computation of the entropy for fermionic zero modes will
require the charges of these fermionic generators under the cartans
of the algebra. Let us choose $\hat{\cal L}_{0}=\frac{L_{1}+L_{-1}}{2}$
as the Cartan of $SL(2,R)$, $J_{3}$ as the Cartan of $SU(2)_{L}$ and
$\hat{J}_{3}$ as the Cartan of $U(1)_{x_{4}}$ (or the $SU(2)_{R}$ for
the non-rotating case). Then
\begin{equation}
\left[\hat{{\cal L}}_{0},G_{n}^{\alpha\beta}\right]  =  -n\tilde{G}_{n}^{\alpha\beta},\quad \left[\hat{J}_{3},G_{n}^{\alpha\beta}\right]  =  \frac{\beta}{2}G_{n}^{\alpha\beta},
\end{equation}
which we also write as
\begin{equation}\label{orbifoldcommutatorg}
    \left[\hat{{\cal L}}_{0}-\hat{J}_{3},G_{n}^{\alpha\beta}\right]=-\left(n+{\beta\over 2}\right)\tilde{G}_{n}^{\alpha\beta},
\end{equation}
and
\begin{equation}
\left[\hat{J}_{3R},G_{n}^{\alpha\beta}\right]  = 0,
\end{equation}
where the last equality is because $U(1)_{x_{4}}$ (or $SU(2)_{R}$ in the non-rotating limit),
commutes with $\hat{su}(1,1|2)$. Hence the $\hat{{\cal L}}_{0},J_{3},\hat{J}_{3}$
charges of $G_{n}^{\alpha\beta}$ are $-n,\frac{\beta}{2},0$ respectively.
\subsection{The Near Horizon Geometry in Different Coordinates.}
The near horizon geometry of the BMPV black hole as presented in \eqref{eq:BMPVOld} does not have AdS$_2$ written out in global coordinates. However, for the computations in the subsequent sections it is useful to have that form. In this section we will provide the transformation of \eqref{eq:BMPVOld} to a coordinate system where the AdS$_2$ is written in global coordinates. 
For this, define the coordinates $\rho$, $\theta$, $\tilde{\chi}$, $\tilde{x}_{4}$ as 
\begin{eqnarray}\label{coordtransform}
\cosh(\eta)=\frac{1}{2}\left(\rho+\rho^{-1}-\rho\tau^{2}\right)\hspace{20mm}e^{-2\theta}=\frac{(1-\tau)^{2}-\rho^{-2}}{(1+\tau)^{2}-\rho^{-2}}\\
\chi=\tilde{\chi}-A\ f(\eta,\theta)\hspace{30mm}x_{4}=\tilde{x}_{4}-B\ f(\eta,\theta)
\end{eqnarray}
where $f(\eta,\theta)\equiv2\tanh^{-1}\left[\tanh[\frac{\theta}{2}]e^{-\eta}\right]$
. In these coordinates the metric becomes 
\begin{eqnarray}\label{NearHorMink}
ds^{2} & = & r_{0}(d\eta^{2}-\sinh^{2}\eta\ d\theta^{2})+r_{0}(d\psi^{2}+\sin^{2}\psi d\phi^{2})+(d\tilde{\chi}-A\cosh\eta d\theta)^{2}\label{eq:BMPVetatheta}\\
 &  & +r_{0}(d\tilde{x}_{4}+\cos\psi d\phi-B\cosh\eta d\theta)^{2}+\nonumber \\
 \nonumber
 &  & +\frac{J}{4r_{0}}(d\tilde{\chi}-A\cosh\eta d\theta)(d\tilde{x}_{4}+\cos\psi d\phi-B\cosh\eta d\theta)
\end{eqnarray}
It will be convenient to redefine $SL(2,R)$ algebra. Let us define
$\hat{L}_{i}$ defined as 
\begin{eqnarray}
\hat{L}_{0} & = & \frac{L_{-}}{2}-L_{+} =\partial_{\theta}\\
\hat{L}_{-} & = & -L_{0}+\frac{L_{-}}{2}+L_{+}=e^{-\theta}(\partial_{\eta}+\coth\eta\partial_{\theta}+ { A \partial_{\chi} + B \partial_{x_4}\over \sinh \eta})\\
\hat{L}_{+} & = & L_{0}+\frac{L_{-}}{2}+L_{+}=e^{\theta}(-\partial_{\eta}+\coth\eta\partial_{\theta}+{ A \partial_{\chi} + B \partial_{x_4}\over \sinh \eta})
\end{eqnarray}
These generators of course obey the $SL(2,R)$ algebra 
\begin{equation}
[\hat{L}_{0},\hat{L}_{\pm}]=\pm\hat{L}_{\pm}\hspace{20mm}[\hat{L}_{+},\hat{L}_{-}]=-2\hat{L}_{0}
\end{equation}

\subsection{Euclidean Near Horizon Geometry}
Since the quantum entropy function is formulated as a Euclidean path integral, it will be convenient 
to have the Euclidean version of the above BMPV geometry. Consider the following analytic continuation
\begin{equation}
    \theta \rightarrow i \theta \hspace{15mm} B \rightarrow -i B \hspace{15mm} A \rightarrow -i A
\end{equation}
Then the metric given in \eqref{NearHorMink} becomes 
\begin{equation}\label{nhgeucl}
\begin{split}
    ds^{2} &=  r_{o} \left[ d\eta^{2}+\sinh^{2}\eta\ d\theta^{2} + d\psi^{2}+\sin^{2}\psi d\phi^{2} + 
(d\tilde{x}_{4}+\cos\psi d\phi-B\cosh\eta d\theta)^{2} \right] \\ & +(d\tilde{\chi}-A\cosh\eta d\theta)^{2} +\frac{J}{4r_{o}}(d\tilde{\chi}-A\cosh\eta d\theta)(d\tilde{x}_{4}+\cos\psi d\phi-B\cosh\eta d\theta).
\end{split}
\end{equation}
Next, taking the limit \eqref{scalinggravity} of \eqref{nhgeucl} gives us the five dimensional effective geometry 
\begin{equation}\label{bmpv}
\begin{split}
ds^2 = r_o\left\lbrace d\eta^2 +\sinh^2\eta\, d\theta^2 + d\psi^2 +\sin^2\psi\, d\phi^2 +\left(1-{J^2\over 64r_o^3}\right)\times\right.\\\left. \left(dx_4 +\cos\psi\,d\phi -B\cosh\eta\,d\theta\right)^2\right\rbrace.
\end{split}
\end{equation}

\section{Constructing Saddle Points of the QEF}\label{sec:qefsaddle}
In this section we shall describe how new saddle points of the quantum entropy function associated with BMPV black holes may be constructed by taking orbifolds of the near horizon geometry. For this, we will use the results of \cite{Banerjee:2009af}, where a class of saddle points that contribute non-vanishingly to the path integral \eqref{qef} were identified for supersymmetric black holes whose horizon carries the symmetry $SU(1,1|2)\times{\cal H}$.

We take the bosonic subgroup $SL(2,R)\times SU(2)$ of the $SU(1,1|2)$ supergroup to be generated by $\hat{L}_{0},\hat{L}_{\pm}$ and $J_1$, $J_2$, $J_3$, satisfying the standard commutation relations
\begin{equation}
    \left[\hat{L}_{0},\hat{L}_{\pm}\right]=\pm\,\hat{L}_{\pm},\quad \left[J_{\ell},J_{m}\right]=i\epsilon^{\ell m n}J_{n}.
\end{equation}
Further, we denote the fermionic generators of $SU(1,1|2)$ by $Q_{\alpha},\tilde{Q}_{\alpha}$ for $\alpha\in1,..4$. 
Their commutation relations are available in section 2 of \cite{Banerjee:2009af}. 
Importantly, one can show that $Q_{1}$ and $\hat{L}_{0}-J_{3}$ form a subgroup of $SU(1,1|2)$, called $H_1$.

It may then be argued that the path integral \eqref{qef}
receives contributions from only those saddle points which are 
invariant under the action of group $H_{1}$ \cite{Banerjee:2009af}. The saddle points we shall construct are $H_{1}$ invariant orbifolds of \eqref{nhgeucl} which asymptote to the full black hole near horizon geometry. If we construct a $Z_{s}$ orbifold
by a $U(1)$ generator ${\cal G}$, the condition that the orbifold
is $H_{1}$ invariant just becomes 
\[
[{\cal G},H_{1}]=0
\]
One can easily check that only $\hat{L}_{0}-J_{3}\subset SU(1,1|2)$
satisfies this property. 

It will turn out that the quotient space constructed by orbifolding the near horizon geometry with $\mathcal{G}=\hat{L}_0-J_3$ will contain fixed points. It is possible to cure these fixed point singularities by passing Ramond-Ramond fluxes through them, thus making them well defined string theory solutions. Alternately, one may choose a $U(1)$ group ${\cal U}\subset{\cal H}$ and define
\begin{equation}\label{eq:OrbifoldGen}
{\cal G}=\hat{L}_{0}-J_{3}+{\cal U}.
\end{equation}
$\cal U$ may then chosen so as to remove the orbifold fixed point. In that case, flux quantization turns out to impose constraints on the charges of the black hole, i.e. the configurations exist only when the corresponding flux quantization conditions are met. 
 
\subsection{Exponentially Suppressed Saddle Points}\label{sec:saddles}
We now carry out a $\mathbb{Z}_s$ orbifold of the near horizon geometry \eqref{nhgeucl}, of the type given 
in \eqref{eq:OrbifoldGen} with the choice 
\begin{equation}\label{genorboperator}
    {\cal U}=\tilde{k}\hat{J}_{3}+{k}u=\tilde{k}\partial_{x_{4}}+{k}\partial_{\chi}.
\end{equation}
This has the following action on the near horizon geometry
\begin{equation}\label{generalorb}
(\theta,\phi,\chi,x_{4})\sim(\theta+\frac{2\pi}{s},\phi-\frac{2\pi}{s},\chi+\frac{2\pi k}{s},x_{4}+\frac{2\pi\tilde{k}}{s})
\end{equation}
When both $k$ and $\tilde{k}$ are zero then the orbifold is generated by $\hat{L}_0-J_3$ and has fixed points.

Next, given the boundary conditions of \eqref{qef}, one has to ensure that the orbifold geometry asymptotes to the near horizon geometry of the black hole.
The metric after orbifolding below along with some trivial relabelling
of coordinates becomes, 
\begin{eqnarray}
ds^{2} & = & r_{o}(d\tilde{\eta}^{2}-\sinh^{2}\tilde{\eta}\ d\tilde{\theta}^{2})+r_{o}(d\psi^{2}+\sin^{2}\psi d\tilde{\phi}^{2})+(d\tilde{\chi}-A\cosh\tilde{\eta}d\tilde{\theta})^{2}\nonumber \\
 &  & +r_{o}(d\tilde{x}_{4}+\cos\psi d\tilde{\phi}-B\cosh\tilde{\eta}d\tilde{\theta})^{2}+\nonumber \\
 &  & +\frac{J}{4r_{o}}(d\tilde{\chi}-A\cosh\tilde{\eta}d\tilde{\theta})(d\tilde{x}_{4}+\cos\psi d\tilde{\phi}-B\cosh\tilde{\eta}d\tilde{\theta})
\end{eqnarray}
with the periodicity condition \eqref{generalorb}. Now make a coordidnate transformation
\begin{eqnarray}
\theta=s\tilde{\theta},\ \ \phi=\tilde{\phi}+(1-s)\tilde{\theta},\ \ \eta=\tilde{\eta}-\log s,\ \ \chi=\tilde{\chi}-k\tilde{\theta},\ \ x_{4}=\tilde{x}_{4}-\tilde{k}\tilde{\theta}
\end{eqnarray}
such that the periodicity becomes 
\begin{eqnarray}
(\theta,\phi,\chi)\equiv(\theta+2\pi,\phi,\chi)\equiv(\theta,\phi+2\pi,\chi)\equiv(\theta,\phi,\chi+2\pi))
\end{eqnarray}
However the metric in these coordinates takes the form
\begin{eqnarray}
ds^{2} & = & r_{o}\left(d\tilde{\eta}^{2}-\sinh^{2}\eta\left(1+\frac{(1-s^{-2})e^{-\eta}}{2\sinh\eta}\right)^{2}d\theta^{2}\right)\nonumber \\
 &  & +r_{o}\left(d\psi^{2}+\sin^{2}\psi(d\phi+d\theta-s^{-1}d\theta)^{2}\right)\\
 &  & +\left[d\chi+ks^{-1}d\theta-A\cosh\eta\left(1+\frac{(1+s^{-2})e^{-\eta}}{2\cosh\eta}\right)d\theta\right]^{2}\nonumber \\
 &  & +r_{o}\left[dx_{4}+\tilde{k}s^{-1}d\theta+\cos\psi(d\phi+d\theta-s^{-1}d\theta)-B\cosh\eta\left(1+\frac{(1+s^{-2})e^{-\eta}}{2\cosh\eta}\right)d\theta\right]^{2}+\nonumber \\
 &  & +\frac{J}{4r_{o}}\left[d\chi+ks^{-1}d\theta-A\cosh\eta\left(1+\frac{(1+s^{-2})e^{-\eta}}{2\cosh\eta}\right)d\theta\right]\times\nonumber \\
 &  & \ \ \ \left[dx_{4}+\tilde{k}s^{-1}d\theta+\cos\psi(d\phi+d\theta-s^{-1}d\theta)-B\cosh\eta\left(1+\frac{(1+s^{-2})e^{-\eta}}{2\cosh\eta}\right)d\theta\right].
\end{eqnarray}
From this we see that as $\eta\rightarrow\infty$, various terms in the
above orbifold geometry (like $d\theta^{2}$ etc) approach that in
the unorbifolded geometry.\footnote{Note that the coefficients of terms like $d\theta d\chi$ do not
approach those in \eqref{eq:BMPVetatheta}. From the point of view
of 2d theory living on $AdS_{2}$, these descend to gauge fields.
However, they do not specify the boundary asymptotics because
the entropy function procedure instructs us to integrate over them.
}
Hence this geometry is an admissible saddle point for the quantum entropy function \eqref{qef}. 

The various possible orbifolds by their $\left(k,\tilde{k}\right)$ values, and the arithmetic constraints imposed on the charges are classified in Table \ref{tab:quantization}. 
\begin{table}
\begin{center}
\begin{tabular}{|c|c|}\hline
  Orbifold labelled by $\left(k,\tilde{k}\right)$ & Constraints on Charges\\ \hline
  $(0,0)$ & no constraints\\ \hline
  $(1,0)$ & $s|n$ and $s|Q_5$\\ \hline
  $(0,1)$ & $s|Q_5$\\  \hline
  $(1,1)$ & $s|n$ and $s|Q_5$\\ \hline
\end{tabular}
\end{center}
\caption {Arithmetic Conditions on Charges arising from Flux Quantization Constraints.} \label{tab:quantization} 
\end {table}
The constraint on $Q_5$ is briefly discussed in Appendix \ref{Flux_Quantization}. The constraint on $n$ follows from orbifold invariance since $n$ is the momentum of string along the circle generated by $u = \partial_\chi$. It is straightforward to see that since these are $\mathbb{Z}_s$ orbifolds of the near horizon geometry, their leading behavior is $e^{S_{Wald}/s}$. We now turn to the next-to-leading terms.
 
\section{Computing The Log Terms}\label{sec:logterms}
In this section we shall discuss the next-to-leading order corrections to the saddle points of the path integral \eqref{qef} obtained by taking the orbifolds \eqref{generalorb} of the near horizon geometry \eqref{bmpv}. As mentioned previously, we are working in the regime \eqref{scalinggravity}. In this scaling limit, the black hole near horizon geometry is characterized by a large length scale $\Lambda^{1/2}$. 

From the discussion in the previous section, it is apparent that to the leading approximation, the horizon degeneracy may be expressed as a sum over various saddle-points of the quantum entropy function.
\begin{equation}\label{dhasym1}
d_{hor} = \sum_{s=1}^{\infty} d_{hor|s},\quad d_{hor|s} \simeq e^{S_{\mathrm{Wald}}/s}.
\end{equation}
We are interested in a more refined computation of the horizon degeneracy, where we retain subleading contributions about each saddle point in the large $\Lambda$ limit. We then have
\begin{equation}\label{dhasym2}
d_{hor|s} \simeq e^{S_{\mathrm{Wald}}/s}\left(\Lambda\right)^{c_s} \quad \Rightarrow \quad \ln d_{hor|s} \simeq {S_{\mathrm{Wald}}\over s} + c_s\ln\Lambda.
\end{equation}
The coefficient $c_s$ is the log term referred to in the Introduction.

We now briefly describe how the log term $c$ may be computed from the quantum entropy function. In particular, since we are working on an odd-dimensional manifold, the near horizon geometry of the BMPV black hole, the log term receives contributions only from the zero mode sector of the kinetic operator. We refer the reader to \cite{Sen:2012cj} and \cite{Bhattacharyya:2012ye} for details about this important fact, merely quoting the final result that
\begin{equation}\label{logtermzeromodes}
\ln\mathcal{Z}=\tfrac12 \sum_{\phi\in\lbrace\,\Phi\,\rbrace}\left(\beta_{\phi}-1\right)n^{\phi}_0\ln\Lambda,%\ln {a},
\end{equation}
where $n^{\phi}_0$ is the number of zero modes of the kinetic operator over the field $\phi$, which may equally well be bosonic or fermionic. Now among the fields of supergravity on AdS$_2 \otimes M$, only vectors, the graviton, and gravitini possess zero modes, which in turn correspond to the discrete series of eigenmodes of the kinetic operator \cite{Banerjee:2010qc,Banerjee:2011jp,Sen:2011ba,Sen:2012cj}. The specific values for $\beta$ for these fields have been computed in $d+2$ dimensions, and are found to be \cite{Sen:2012cj}
\begin{equation}\label{betavalues}
\beta_{v}={d\over 2},\quad \beta_{m}={d+2\over 2},\quad \beta_{f}=d+1.
\end{equation}
Here the subscripts $v$ denote the vector field, $m$ the metric or the graviton, and $f$ the gravitino. Hence we see that the computation of the log term for odd-dimensional manifolds reduces to the counting of zero modes in the spectrum of the kinetic operator. Zero modes $f^{(i)}$ can in principle be readily counted by evaluating the expression  
\begin{equation}\label{nzero}
n_0 = \left(-\tfrac12\right)^F\,\sum_{i\in \mathcal{I}_0}\int_{\mathcal{M}} d^{d+2}x \,\sqrt{g}\, f^{(i)*}\l(x\r) \cdot f^{(i)}\l(x\r),
\end{equation}
where $\mathcal{I}_0$ is the index set for the zero modes $f^{(i)}$ and `$\cdot$' is the invariant inner product defined for the wave functions $f$. For instance, if $f^{(i)}$ are vectors over $\mathcal{M}$, then $ f^{(i)*}\cdot f^{(i)} = g^{MN} f^{(i)*}_{M} \cdot f^{(i)}_N$. $F$ is the fermion number, which is $0$ for bosons and $1$ for fermions. 
\subsection{Counting Zero Modes in Exponentially Suppressed Saddle Points}
We now take the BMPV near horizon geometry written in Euclidean signature \eqref{bmpv} and implement the orbifold \eqref{generalorb} on this geometry. Note that the translation along the $\chi$ direction is purely internal in this limit. We will refer to the resulting orbifold spaces as $\mathsf{bmpv}/s$. Finally, the index $M$ runs over the directions $\eta,\,\theta,\,\psi,\,\phi,\,x_4$.
We now compute the number of zero modes of the vector, graviton and gravitini after this orbifold is imposed.
\subsubsection{The Vector Field Zero Modes}
The zero modes of the five-dimensional vector field in the BMPV near-horizon geometry have been enumerated in \cite{Sen:2012cj}. They are the discrete modes of the vector field along AdS$_2$, carrying no support along the squashed S$^3$ directions. In particular,
\begin{equation}\label{vectorbmpv}
\mathcal{A}_M^{(m)}={1\over \mathcal{N}_m}\nabla_M\phi^m,\quad \phi^m={\sqrt{1\over 2\pi\vert m\vert}}\left[\sinh\eta\over 1+\cosh\eta\right]^{\vert m\vert}e^{i m\theta},
\end{equation}
which is the same as \eqref{vectordiscrete}, upto a normalization constant $\mathcal{N}_{m}$, only now the covariant derivative is now with respect to the background metric \eqref{bmpv}. Additionally, invariance under the orbifold \eqref{generalorb} changes the quantization condition from $m\in \mathbb{Z}-\lbrace0\rbrace$, applicable on the unquotiented space, to
\begin{equation}
m= sp, \quad p\in\mathbb{Z}-\left\lbrace 0\right\rbrace.
\end{equation}
We then obtain,
\begin{equation}
g^{MN}\mathcal{A}^{(m)*}_{M}\mathcal{A}^{(m)}_N ={1\over\mathcal{N}_m^2}\frac{\left| m\right| \cosech^2\eta \tanh ^{2 \left| m\right| }\left(\frac{\eta }{2}\right)}{\pi  r_o}.
\end{equation}
The normalization constant $\mathcal{N}_m$ is fixed by requiring that 
\begin{equation}
\int_{\mathsf{bmpv}/s}\,d^2x\,d^3y\sqrt{g} g^{MN}\mathcal{A}^*_{M}\mathcal{A}_N =1.
\end{equation}
Thus, given the normalized set of vector zero modes on $\mathsf{bmpv}/s$, we may use \eqref{nzero} to evaluate the number of zero modes. Naively the answer is divergent, but we regulate the divergence by placing a radial cutoff $\eta_0$ on the AdS$_2$ factor. We obtain
\begin{equation}\label{vbmpvn}
\begin{split}
n_0 &=\sum_{p\in \mathbb{Z}-\lbrace 0\rbrace}{1\over \mathcal{N}^2_{sp}}\int_{\mathsf{bmpv}/s}^{\eta_0} d^5x\sqrt{g} g^{MN} \mathcal{A}^{sp *}_M \mathcal{A}_N^{sp} \\&=\sum_{p\in \mathbb{Z}-\lbrace 0\rbrace}(\tanh\frac{\eta_0}{2})^{2s|p|}\simeq {1\over 2}e^{\eta_0}-1+\mathcal{O}\l(e^{-\eta_0}\r).
\end{split}
\end{equation}
We drop the factor diverging with the AdS radial coordinate, and keep the order 1 term as the number of zero modes. Hence the number of zero modes from a vector field on the five dimensional space $\mathsf{bmpv}/s$ is given by
\begin{equation}\label{vzerobmpvn}
n_0^{\mathcal{A}}=-1.
\end{equation}
\subsubsection{The Graviton Zero Modes}
We next turn to the graviton zero modes, for which the number of zero modes is obtained by applying \eqref{nzero} to find
\begin{equation}\label{gwbmpvn1}
n_0=\sum_{\ell}{1\over \mathcal{N}^2_{\ell}}\int_{\mathsf{bmpv}/s} d^2x\,d^3y\,\sqrt{g} g^{MP}g^{NQ} w^{\ell\,*}_{MN}w^{\ell}_{PQ}.
\end{equation}
Now the zero modes of the graviton come in two sets \cite{Sen:2012cj}. Firstly we have the modes enumerated in \eqref{gravitonbmpv} below, which obey the quantization condition $\ell=sp$, where $p\geq 1$ for invariance under the orbifold \eqref{generalorb}. We denote these modes as $w^{(0)}$.
\begin{equation}\label{gravitonbmpv}
w^{(0)\ell}_{\mu\nu}= {1\over \mathcal{N}_\ell} h^{\ell}_{\mu\nu},\quad w^{(0)\ell}_{\mu a} =0,\quad w^{(0)\ell}_{ab}=0.
\end{equation}
where $h^{\ell}$ has been defined in \eqref{metricdiscrete}. Further, they now have to be normalized over the quotient space $\mathsf{bmpv}/s$. The normalization constant $N_\ell$ is determined through
\begin{equation}\label{gravwnorm1}
{1\over \mathcal{N}^2_\ell}\int_{\mathsf{bmpv}/s} d^2x\,d^3y\,\sqrt{g} g^{MP}g^{NQ} w^{(0)\ell\,*}_{MN}w^{(0)\ell}_{PQ}=1,
\end{equation}
With this normalization of the zero modes \eqref{gravitonbmpv}, we use \eqref{gwbmpvn1} to count the number of zero modes, regulating the divergence in $n_0$ by placing a cutoff $\eta_0$ on the AdS$_2$ radial coordinate. 
We eventually find that the number of zero modes is given by
\begin{equation}
\begin{split}
n_{0}^{w(0)}&=\sum_{p=1}^\infty\tanh^{2sp}\l({\eta_0\over 2}\r)\l(2+4sp{\cosh\eta_0\over\sinh^2\eta_0}+4(sp)^2{1\over\sinh^2\eta_0}\r)\\& \simeq {3\over 2s}e^{\eta_0}-1+\mathcal{O}\l(e^{-\eta_0}\r).
\end{split}
\end{equation}
Hence this contribution to the graviton zero modes is given by
\begin{equation}\label{nzerow0}
n^{w(0)}_0=-1.
\end{equation}
The next class of graviton zero modes are obtained by taking the tensor product of discrete modes along AdS$_2$ and Killing vectors along squashed S$^3$. Expressions for the Killing vectors $J_i$ of squashed $S^3$ have already been given in \eqref{bmpvkilling1} but the basis $k^{(i)}$ presented in \eqref{bmpvkilling} is more convenient for our purposes here. We will therefore consider the set of discrete modes $w^{(i)}$ of the five-dimensional graviton given by
\begin{equation}
w^{m(i)}_{\mu a} = {1\over\mathcal{N}_{m(i)}}\mathcal{A}^m_{\mu}k^{(i)}_{a} = w^{m(i)}_{a \mu}, \quad i = 1,2,3,4,
\end{equation}
with all other components of $w^{(i)}$ being zero. The zero modes $w^{m(i)}$ have the following $\left(\theta,\phi\right)$ dependence, 
\begin{equation}
w^{m(1)} \sim e^{i m\theta}e^{-i\phi},\quad w^{m(2)} \sim e^{i m\theta}e^{i\phi},\quad w^{m(3)} \sim e^{i m\theta}, \quad w^{m(4)} \sim e^{i m\theta},
\end{equation}
and are independent of $x_4$. Hence we see that the orbifold invariant modes satisfy
\begin{equation}\label{allowedmbmpv}
\begin{split}
w^{m(1)}\quad &:\quad m=sp-1,\quad p\in\mathbb{Z}-\lbrace 0\rbrace,\\
w^{m(2)}\quad &:\quad m=sp+1,\quad p\in\mathbb{Z}-\lbrace 0\rbrace,\\
w^{m(3)}\quad &:\quad m=sp,\quad p\in\mathbb{Z}-\lbrace 0\rbrace,\\
w^{m(4)}\quad &:\quad m=sp,\quad p\in\mathbb{Z}-\lbrace 0\rbrace.\\
\end{split}
\end{equation}
We also have to normalize the zero modes appropriately over the orbifold space $\mathsf{bmpv}/s$. The procedure for doing this has already been described above, and we shall only describe final results. 
It turns out that in all four cases we have to compute the sum,
\begin{equation}
n_0^{w;i} = \sum_{m} \tanh ^{2 \left| m\right| }\left(\frac{\eta _0}{2}\right)
\end{equation}
over the values of $m$ enumerated in \eqref{allowedmbmpv}. On doing so, and retaining the order 1 term in the large $\eta_0$ expansion as per our usual prescription, we finally obtain that
\begin{equation}\label{nzerowi}
n_0^{w;1} = 0,\quad n_0^{w;2} = 0,\quad n_0^{w;3} = -1,\quad n_0^{w;4} = -1.
\end{equation}
Then the total number of graviton zero modes is given by adding \eqref{nzerow0} and \eqref{nzerowi}. We finally obtain for the total number of graviton zero modes on $\mathsf{bmpv}/s$,
\begin{equation}
n_0^w = -3.
\end{equation}
\subsubsection{Counting Gravitini Zero Modes}
Gravitini zero modes are associated with the deformations generated by the fermionic generators $G_{n}^{\alpha\beta}$ of the $\mathcal{N}=4$ superconformal algebra, where $\alpha,\,\beta=\pm 1$, and $n=\mathbb{Z}+{1\over 2}$ \cite{Sen:2009vz,Banerjee:2009af}. Of these, $G_{\pm \tfrac12}^{\alpha\beta}$ correspond to global symmetry generators in the algebra $su(1,1|2)$ of the near horizon geometry. The remaining generators may be identified to the discrete modes $\xi_{m}^{(k)+}$ and $\xi_{m}^{(k)+}$ along AdS$_2$, enumerated in Appendix \ref{ads2app}, in the following manner. Firstly, the $G_{n}^{\alpha\beta}$ for $n\geq \frac32$ are identified to the modes $\xi_{m}^{(k)+}\otimes\Psi$ where $n=k+\frac12$ and $\Psi$ is a spinor along the compact directions. Next, the generators $G_{n}^{\alpha\beta}$ for $n\leq -\frac32$ correspond to the modes $\hat{\xi}_{m}^{(k)+}\otimes\Psi$ where $\vert n\vert=k+\frac12$. 

To determine the zero modes that survive the orbifold projection, we will use the expression \eqref{orbifoldcommutatorg}, which may be exponentiated to yield the orbifold action on the fermion zero modes
\begin{equation}
e^{-2\frac{\pi}{s} i\left(\mathcal{L}_0-\hat{J}^3\right)}G_{n}^{\alpha\beta}e^{2\frac{\pi}{s} i\left(\mathcal{L}_0-\hat{J}^3\right)} =e^{-2\frac{\pi}{s} i\left(n+\frac\beta2\right)}G_{n}^{\alpha\beta}.
\end{equation}
The orbifold invariant modes are given by the solutions in $n$ of the equation $n+{\beta\over 2} = sp$, where $p\in\mathbb{Z}$. These have been enumerated in Table \ref{tab:title}.
\begin{table}
\begin{center}
\begin{tabular}{|c|c|c|}\hline
  Range of  $n$ & $\beta$ & Solution\\ \hline
  $\quad n\geq\frac32 \Leftrightarrow k\geq 1,\quad$ &$\quad +1 \quad$ & $\quad k=sp-1\quad$\\
  $n = k+\frac12$ &$\quad -1 \quad$ & $\quad k=sp\quad$\\ \hline
    $\quad n\leq\frac32 \Leftrightarrow k\geq 1,\quad$ &$\quad +1 \quad$ & $\quad k=sp\quad$\\
  $n = -k-\frac12$ &$\quad -1 \quad$ & $\quad k=sp-1\quad$\\ \hline
\end{tabular}
\end{center}
\caption {Quantization Conditions on the Fermion Zero Modes imposed by Orbifold Invariance.} \label{tab:title} 
\end {table}
Note that the value of $\alpha$ is not constrained in the above projection. Now we proceed to the counting of fermionic zero modes. As noted, they appear in the discrete series of gravitini modes and correspond to the following configurations.
\begin{equation}
\Xi_{\eta}^{(k)+} = {C\over 4\pi a}\,
\xi_{\eta}^{(k)+}\otimes\Psi,\quad \Xi_{\theta}^{(k)+} = {C\over 4\pi a}\,\xi_{\theta}^{(k)+} \otimes\Psi,
\end{equation}
and the hatted spinors are given by
\begin{equation}
\hat{\Xi}_{\eta}^{(k)+} = {C\over 4\pi a}\,\hat{\xi}_{\eta}^{(k)+}\otimes\Psi,\quad \hat{\Xi}_{\theta}^{(k)+} = {C\over 4\pi a}\,\hat{\xi}_{\theta}^{(k)+}\otimes\Psi.
\end{equation}
The normalization constant $C$ is fixed by demanding that the spin-$\frac32$ fields $\Xi^{+}_{\mu}$ and $\hat{\Xi}^{+}_{\mu}$ are Kronecker delta function normalized over the space $\mathsf{bmpv}/s$.
Now with this choice of normalization we can compute explicitly and show that
\begin{equation}
\bar{g}^{mn}\left(\xi^{k}_{m}\right)^\dagger \xi^{k}_{n} = \bar{g}^{mn}\left(\hat{\xi}^{k}_{m}\right)^\dagger \hat{\xi}^{k}_{n} ={\sinh^{2k-2}{\eta\over 2}\over\cosh^{2k+4}{\eta\over 2}}.
\end{equation}
The contribution to the number of zero modes from the series $\xi\otimes\Psi$ may now be obtained by using \eqref{nzero}. We obtain
\begin{equation}
\begin{split}
n^0_{\xi} = {-\frac12}\cdot 2&\sum_{p=1}^\infty \left(\int_{0}^{\infty}d\eta\, \sinh\eta\,\bar{g}^{mn}\left(\xi^{sp}_{m}\right)^\dagger \xi^{sp}_{n}\right.\\&\qquad\qquad +\left. \int_{0}^{\infty}d\eta\, \sinh\eta\, \bar{g}^{mn}\left(\xi^{sp-1}_{m}\right)^\dagger \xi^{sp-1}_{n}\right).    
\end{split}
\end{equation}
The overall factor of $2$ is because of the multiplicity of zero modes associated with $\alpha=\pm 1$. As it stands, the above expression is divergent, but may be regulated to obtain
and we regulate it by writing
\begin{equation}
n^0_{\xi} = -\left(\frac{e^{\eta _0}}{s}-1 +\mathcal{O}\left(e^{-\eta_0}\right)\right).
\end{equation}
An entirely analogous procedure may be applied to the hatted spinors, and we obtain
\begin{equation}
n^0_{\xi} = +1,\quad n^0_{\hat{\xi}} = +1.
\end{equation}
Therefore, the total number of zero modes is
\begin{equation}
n^0_{f} = +2.
\end{equation}
\subsection{Counting Zero Modes for the Non-Rotating Black Hole}\label{nonrotatingzm}
As mentioned in Section \ref{nonrotatingnhg}, in the limit where $J=0$, the near horizon geometry factors into the tensor product AdS$_2\otimes S^3$ and we obtain two additional Killing vectors, given in \eqref{extrakilling}, which lead to extra zero modes of the five-dimensional graviton, given by
\begin{equation}
w^{m(i)}_{\mu a} = {1\over\mathcal{N}_{m(i)}}\mathcal{A}^m_{\mu}\tilde{k}^{(i)}_{a} = w^{m(i)}_{a \mu}, \quad i = 1,2.
\end{equation}
It turns out that these graviton zero modes make the log term sensitive to the choice of orbifold, as we shall now describe. Firstly we have the Type $(0,0)$ and Type $(1,0)$ orbifolds, which act on the five-dimensional geometry as 
\begin{equation}\label{orb1}
\left(\theta,\phi,x_4\right) \mapsto \left(\theta+{2\pi\over s},\phi-{2\pi\over s},x_4\right).
\end{equation}
In this case the orbifold-invariant modes of $\tilde{w}^{m(1)}$ and $\tilde{w}^{m(2)}$ satisfy the quantization condition $m=Np$, and contribute $-1$ each to the number of metric zero modes.  In contrast the Type $(0,1)$ and Type $(1,1)$ orbifolds act as
\begin{equation}\label{orb2}
\left(\theta,\phi,x_4\right) \mapsto \left(\theta+{2\pi\over s},\phi-{2\pi\over s},x_4+{4\pi\over s}\right).
\end{equation}
In this case, the orbifold-invariant modes of $\tilde{w}^{m(1)}$ and $\tilde{w}^{m(2)}$ satisfy the quantization condition $m=sp-2$ and $m=sp+2$ respectively. From the methods outlined previously, it may readily be seen that both $\tilde{w}_1$ and $\tilde{w}_2$ contribute zero to the regularized number of discrete modes.
\subsection{The Log Terms for the BMPV Black Hole}
We may now put together the results of the zero mode counting with the equations \eqref{logtermzeromodes} and \eqref{betavalues} to compute the log term about each exponentially suppressed saddle point \footnote{The Type $(0,0)$ orbifolds have fixed point singularities which may be resolved by passing Ramond Ramond fluxes through them. While it is possible that there are new states localized on these fixed points corresponding to twisted sectors of the string, but the contribution of these states to the partition function would not scale with $\Lambda$. Hence we expect our computation to hold for these orbifolds as well.}. It is straightforward to obtain that if 
the effective five dimensional theory has $n_V$ massless $U(1)$ gauge fields,
\begin{equation}\label{rotatinglog}
\begin{split}
\ln\mathcal{Z}_{1-\ell}&={1\over 2}\left[n_V\left({3\over 2}-1\right)\left(-1\right) + \left({5\over 2}-1\right)\left(-3\right) +\left(4-1\right)2\right]\ln \Lambda\\
&=-{1\over 4}\left(n_V-3\right)\ln\Lambda.
\end{split}
\end{equation}
This is the same answer as arrived at over the dominant saddle point \cite{Sen:2012cj} despite the fact that the numbers of regularized zero modes for each fields change separately. Hence, the quantum entropy function predicts that the value of the log term should be independent of the choice of saddle point. This is reminiscent of the result for large black holes in $\mathcal{N}=4$ and $\mathcal{N}=8$ string compactifications in four dimensions \cite{Gupta:2014hxa}.

For comparison with the microscopic side, we shall focus on the compactification where $\mathcal{M}=T^5$, and hence $n_V=27$. For this case we find
\begin{equation}\label{rotatinglogt5}
\ln\mathcal{Z}_{1-\ell}=-6\,\ln\Lambda.
\end{equation}
\subsubsection{The Non-Rotating Case}
As we have seen in Section \ref{nonrotatingzm}, in this case the number of orbifold invariant zero modes, and hence the log term, becomes sensitive to the choice of orbifold. For this reason we will organize the answers for different orbifolds \eqref{generalorb} according to the arithmetic conditions they obey.
\paragraph{No Arithmetic Constraints:} These are the Type (0,0) Orbifolds of Section \ref{sec:saddles} and using the zero mode counting of Section \ref{nonrotatingzm}, we obtain that the log term about these saddle points is given by
\begin{equation}\label{logt00t10}
\ln\mathcal{Z}_{1-\ell} = -{n_v+3\over 4}\ln\Lambda.
\end{equation}
For the case of $\mathcal{M}=T^5$, we therefore find
\begin{equation}\label{nonrotatinglogt5:00}
\ln\mathcal{Z}_{1-\ell}=-\frac{15}{2}\,\ln\Lambda.
\end{equation}
\paragraph{Constraint that $s|Q_5$:} These are the Type (0,1) Orbifolds, and again using the zero mode counting of Section \ref{nonrotatingzm}, we see that the log term about these saddle points is given by
\begin{equation}\label{logt01t11}
\ln\mathcal{Z}_{1-\ell} = -{n_v-3\over 4}\ln\Lambda.
\end{equation}
\paragraph{Constraint that $s|n$ and $s|Q_5$:} These are the Type (1,0) and Type (1,1) Orbifolds. Now the log term about the Type (1,0) orbifold is given by \eqref{logt00t10} and about the Type (1,1) Orbifold is given by \eqref{logt01t11}.
Thus, the contribution of the (1,1) Orbifold to the path integral dominates over the contribution of the (1,0) Orbifold, and we write
\begin{equation}\label{nonrotatinglogt5:constrained}
\ln\mathcal{Z}_{1-\ell} = -{n_v-3\over 4}\ln\Lambda.
\end{equation}
In particular, for the Type (0,1), (1,0) and (1,1) orbifolds for the case where $\mathcal{M}=T^5$, we obtain
\begin{equation}\label{nonrotatinglogt5}
\ln\mathcal{Z}_{1-\ell} = -{6}\ln\Lambda.
\end{equation}
\section{Large Charge Expansion in Type II String on T$^{5}$}\label{sec:microt5}
This section is a review of some essential facts about the microscopic computation of the BMPV black hole entropy in the toroidally compactified Type II string that will be useful when making a comparison to macroscopic results. The presentation is by no means exhaustive and for more details we refer the reader to \cite{Dabholkar:2010rm,Maldacena:1999bp} where the microscopic analysis is available. The computation of the log term about the dominant saddle point is available in \cite{Sen:2012cj}. The microscopic system at hand consists of $Q_5$ D5-branes wrapped on $T^{5}=T^{4}\times S^{1}$, along with $Q_1$ units of D1-brane charge wrapped along $S^1$, $-n$ units of momentum along the $S^1$, as well as $J_{3L}=J/2$ units of $SU(2)_L$ angular momentum. The computation of microscopic degeneracy proceeds via the computation of an appropriate index over this system. However, the choice of index depends crucially upon whether $J$ is zero or non-zero and hence we shall treat these two cases separately.
\subsection{The Rotating Case}
When $J$ is non-zero the D1-D5 system on $T^{5}=T^{4}\times S^{1}$describes a 1/8 BPS
state. Defining $Q\equiv Q_{1}Q_{5},$ the index is given by \cite{Dabholkar:2010rm,Maldacena:1999bp}
\begin{equation}
\sum_{J}(-1)^{J}\tilde{d}_{micro}(n,Q,J)e^{2\pi iJv}=(e^{i\pi v}-e^{-i\pi v})^{4}\sum_{j\in\mathbb{Z}}\ \sum_{\s|n,Q,j}\ \s\hat{c}\left(\frac{4Qn-j^{2}}{\s^{2}}\right)e^{2\pi ijv}.\label{eq:defdT5}
\end{equation}
Here the notation $a|b$ denotes that $a$ is a divisor of $b$. Further , the function $\hat{c}$ is defined via the Fourier coefficients of the modular form
\begin{equation}
-\frac{\theta_{1}(v,\tau)^{2}}{\eta(\tau)^{6}}=\sum_{k,\ell}\hat{c}(4k-\ell^{2})e^{2\pi i(k\tau+\ell v)},\quad k,\ell\in\mathbb{Z}\label{eq:defc}
\end{equation}
where $\theta_{1}(v,\tau)$ is a Jacobi Elliptic Theta Function and and $\eta(\tau)$ is the Dedekind eta function. Since $\s|j$ and $\s^{2}|Qn$ in \eqref{eq:defdT5}, the argument of $\hat{c}$ in the RHS of \eqref{eq:defdT5} is exactly of the form given in \eqref{eq:defc}. 
We may solve \eqref{eq:defdT5} for $\tilde{d}_{micro}$ to obtain \begin{equation}
\label{MainEqn1}
\tilde{d}_{micro}\left( n,Q,J \right) = \left(-1\right)^J \sum_{q=-2}^{2} \lambda_q  \sum_{\s|n,Q,J+q} \bar s\,\hat{c}\left({4Qn-(J+q)^2\over \bar s^2}\right), 
\end{equation}
where we have defined the constants $\lambda_0 = 6, \lambda_{\pm 1}=-4, \lambda_{\pm 2}=1$. 
Note that this leads to an arithmetic constraint on $J$ that dictates which values of $\bar s$ may enter the equation \eqref{eq:defdT5} for a given $J$. In particular, only those values of $\s$ are allowed such that at least one of $J,J \pm 1, J \pm 2$ is $p \s$ where $p \in {\mathbb Z}_+$.
Let us now consider a particular term above with $s$ satisfying $\bar s|Q, \bar s| n$ 
\begin{equation} \label{eachstorus}
\tilde{d}_{micro}\left( n,Q,J \right)|_{\s} = (-1)^J \s \sum_{q=-2 , \bar s | J +q}^2 \lambda_q \hat c\left(D_q \right) 
\end{equation}
where we have defined
\begin{equation}
D_q \equiv 4 \left({Q \over \s}\right) \left({n \over \s}\right) - \left({J+q \over \s}\right)^2
\end{equation}
The large charge behavior of the $\hat{c}$ is dictated by the Rademacher expansion of these Fourier coefficients, worked out for instance in \cite{Dabholkar:2011ec} to which we refer the reader for background and details. 
We then find that the Fourier coefficients take the form
\begin{equation}\label{rademacher}
    \hat{c}(D) =  \sum_{c=1}^\infty {\sqrt 2 K_c(D) \over \sqrt c D^2} e^{\pi \sqrt D \over c} \left[ 1 -   {6 c \over \pi \sqrt D } + {\cal O}(1/D)\right] \hspace{10mm} \mbox{for large }D
\end{equation}
where $K_c(D)$ is the  Kloosterman sum. For low values of $c$, one may readily evaluate it to obtain
\begin{equation}\label{kloosterman}
K_1(D) = {(-1)^{D \over 2} \over \sqrt 2},\quad K_2(D) = \begin{cases} 0 & \mbox{ if $D$ is odd}\\
e^{- i D \pi \over 4} & \mbox{ if $D$ is even}
\end{cases}.
\end{equation}
For each term \eqref{eachstorus} we get, we get 
\begin{equation}\label{eachstorusnew}
\tilde{d}_{micro}\left( n,Q,J \right)|_{\s} = (-1)^J \sqrt 2  \bar s \sum_{q=-2, \bar s | J+q}^2 \lambda_q \sum_{c=1}^\infty {  K_c(D_q) e^{\pi \sqrt D_q \over c } \over \sqrt c D_q^2} \left( 1 - {6 c \over \pi \sqrt{D_q}  } + {\cal O}({1 \over D_q}) \right).
\end{equation} 
Let us now define a new charge 
\begin{equation}
D \equiv   4 n Q - J^2 
\end{equation}
in terms of which we will write all the subsequent expressions. Since ${q(q+2J) \over D} \ll 1$,  \eqref{eachstorusnew} now becomes,
\begin{equation}\label{sumeachtorus}
\begin{split}
   &\tilde{d}_{micro}\left( n,Q,J \right)|_{\s} = {\sqrt{2} (-1)^J \bar s^5 \over D^2}  
\sum_{c=1}^\infty  {e^{\pi \sqrt D  \over c \bar s} \over \sqrt c}  \left(1 - {6 c \sqrt{\bar s} \over \pi \sqrt D } + {\cal O}({1 \over D})\right) \times\\
&\sum_{q=-2, \bar s | J+q}^2 \lambda_q  K_c(D_q)  e^{-{ \pi q (q+2J) \over 2 c \bar s \sqrt{D} }}  \left( 1+{\cal O}({J \over D^{3 \over 2}}) \right) 
 \left(1 + {2 q (q+2J) \over D}  + {\cal O}({J \over D^2 }) \right).
\end{split}
\end{equation}
Focusing on the rotating case, where $J \sim \sqrt D$, we get 
\begin{eqnarray}\label{sumeachtorusrot}
\tilde{d}_{micro}\left( n,Q,J \right)|_{\s} &=&  {\sqrt{2} (-1)^J \bar s^5 \over D^2}  
\sum_{c=1}^\infty  {e^{\pi \sqrt D  \over c \bar s} \over \sqrt c}  \left(1 - {6 c \sqrt{\bar s} \over \pi \sqrt D } \right) \\ 
\nonumber
&& \hspace{10mm} \times \sum_{q=-2, \bar s | J+q}^2 \lambda_q  K_c(D_q) e^{-{ \pi q J \over  c \bar s \sqrt{D} }}
 \left(1 + {4 q J \over D}  + {\cal O}({1 \over D }) \right).
\end{eqnarray}
\textbf{$\s>4$ case:} For simplicity, consider the case $\bar s >4$ and thus $\bar s$ can atmost divide only one of the integers in $J-2 ,J-1,,J+2$. In this case, we see that only one of the terms in the sum over $q$ is nonvanishing.  Hence the scaling is  .
\begin{equation}
\tilde{d}_{micro}\left( n,Q,J \right)|_{\s}  \sim \sum_{c=1}^\infty {e^{\pi \sqrt{D} \over c \bar s} \over D^2}
\end{equation}
Hence we find that for all $\s > 4$, (upto quantization conditions on $n,Q,J$) we have 
\begin{equation}\label{dmicrosa}
\tilde{d}_{micro}\left(n,Q,J\right)\vert_{\s}\simeq\sum_{c=1}^\infty e^{\pi\sqrt{4Qn-J^2}/c\s}\left(4Qn-J^2\right)^{-2}.
\end{equation}
We therefore have a series of exponentially suppressed corrections to the leading `Bekenstein-Hawking' contribution for each $\s$. In particular
\begin{equation}
\log \tilde{d}_{micro}\left(n,Q,J\right)|{(\s,c)} \simeq {\pi\sqrt{4Qn-J^2}\over c\s} -2\log\left(4Qn-J^2\right).
\end{equation}
Hence under the scaling \eqref{scalinggravity}
of charges, the log term about each saddle point is $-6\ln\Lambda$.\\
\textbf{Comments on $\bar s \leq 4$ case: } In this case it could be that $\bar s$ divides more than one integer in the set $(J-2,..J+2)$. Then more than one term is nonvanishing in the sum in \eqref{sumeachtorus}. Generically, there are no cancellations among these terms and the result is the same as $\bar s>4$ case. 
Nonetheless, in principle it is possible that delicate cancellations between individial terms might alter the log correction \footnote{This possibility is not merely hypothetical. It may be shown that the change in the log term about the dominant saddle point from $-6\ln\Lambda$ for the rotating case to $-\tfrac{15}{2}\ln\Lambda$ for the non-rotating case \cite{Sen:2012cj} may be traced back to precisely this cancellation between different terms that contribute to the index.}.  The simplest case this might happen is for $\bar s = 4$, and $J + 2$ (and hence $J-2$ as well) is 
divisible by $\bar s$, i.e $q = \pm 2$ terms are nonvanishing in the sum \eqref{sumeachtorus}. In this case, the log correction will be different from $\bar s>4$ case if 
\begin{eqnarray}  
K_c(D_{-2}) e^{-2\pi J \over c \bar s \sqrt D} + K_c(D_{2}) e^{2\pi J \over c \bar s \sqrt D}   = 0
\end{eqnarray} 
In practice, such cancellations do not occur when we compute explicitly, for example for the cases of $c=1$ and $c=2$, but it would be interesting to study this question more systematically using properties of the Kloosterman sum.  For now, we will assume that such delicate cancellations do not occur.

\subsection{The Non-Rotating Case}
For the non-rotating case, the relevant microscopic index is $d_{micro}$, given by \cite{Sen:2012cj}
\begin{equation}
    d_{micro}(n,Q,J) = \tilde{d}_{micro}\left(n,Q,J\right) - \tilde{d}_{micro}\left(n,Q,J+2\right),
\end{equation}
which, for $J=0$ takes the form
\begin{eqnarray}\nonumber
d_{micro}(n,Q,0) &=& \tilde{d}_{micro}\left(n,Q,0\right) - \tilde{d}_{micro}\left(n,Q,2\right) \\
\nonumber
&& =  \sum_{j\in\mathbb{Z}}\sum_{\bar s|n,Q,j} \bar s\, \hat{c}\left({4Qn-j^2\over \bar s^2}\right)    \sum_{q=-2}^4 \tilde  \lambda_q \delta_{j,q},
\end{eqnarray}
where $\tilde \lambda_{-2} = 1,\tilde \lambda_{-1} =- 4, \tilde \lambda_{0} = 5,\tilde \lambda_{1} = 0,\tilde \lambda_{2} = -5, \tilde \lambda_{3} = 4, \tilde \lambda_{4} = -1$. This is just of the form  \eqref{MainEqn1} and hence we can read off the answer in \eqref{sumeachtorus}, now with $J=0$ to obtain 
\begin{eqnarray}
d_{micro}(n,Q,0) = \sum_{q=-2}^4 \tilde \lambda_q \sum_{\bar s |n, Q, q} {\sqrt{2} \bar s^5 \over D^2} \sum_{c=1}^\infty { e^{\pi \sqrt D \over c \bar s} \over \sqrt c} K_c({D - q^2 \over \bar s^2}) \left[ 1 - { \left( {\pi q^2 \over 2 c \bar s } + {6 c \sqrt{\bar s} \over \pi} \right)  \over \sqrt D} \right],
\end{eqnarray}
where we have dropped terms  ${\cal O}({1 \over D})$. Again one can make a similar argument as in $J \ne 0$ case to say that atleast for $\bar s > 4$, only the $q=0$ term in the sum will contribute. Hence  atleast for $\bar s >4$, the log corrections are unchanged and continue to be given by $-6\ln\Lambda$.  

When $\bar{s}<4$, as in the rotating case, we may investigate the possibility of cancellations between different contributing terms for some low values of $c$ which we turn to below. The values of the Kloosterman sum relevant for these computations have already been detailed in \eqref{kloosterman}.
\paragraph{The $c=1$ case:} This includes the case of the dominant saddle point, which corresponds to $\bar{s}=1,\,c=1$.  In this case, to leading order in $D$, 
\begin{eqnarray}\nonumber
d_{micro}(n,Q,0)|_{c=1,\bar s|n,Q} &=&  {\sqrt{2} \bar s^5 (-1)^{D \over 2 \bar s^2} \over D^2}  e^{\pi \sqrt D \over  \bar s} \left[  5 -(-1)^{-{8 \over  \bar s^2}} \delta_{\bar s | 4}     \right. \\ 
&& \left. \ \ \  + 4 \left( (-1)^{-{9 \over 2 \bar s^2}} \delta_{\bar s | 3} - (-1)^{-{2 \over  \bar s^2}} \delta_{\bar s | 2}-  (-1)^{{1 \over 2 \bar s^2}} \delta_{\bar s , 1}    \right)    \right] .
\end{eqnarray}
Although for $\bar s =1$ this is vanishing, which is the reason for the change in the log term to $-\tfrac{15}{2}\ln\Lambda$, one can check explicitly that this is non zero for any value of $\bar{s} >1$. Therefore, for $c=1$ and $\bar{s}>1$, the log term will be given by $-6\ln\Lambda$.
\paragraph{The $c=2$ case:} Also note that the quantization conditions imply that  ${D \over \bar s^2}$ is always divisible by $4$. With these inputs, one can compute that the contribution to $d_{micro}$ to leading order in $D$ gives
\begin{eqnarray}\nonumber
d_{micro}(n,Q,0)|_{c=2,\bar s|n,Q} &=& {\sqrt{2} \bar s^5 e^{\pi \sqrt D \over  2\bar s}  \over D^2}  \left( 5 K_2({D \over \bar s^2}) - 4 \delta_{\bar s  | 2} K_2({D -4 \over \bar s^2}) - \delta_{\bar s  | 4} K_2({D-16 \over \bar s^2}) \right).
\end{eqnarray}
One can again check that this is nonvanishing for any $\bar s \geq 1$, and hence the log term is again given by $-6\ln\Lambda$.
\section{Summary}
In this paper we examined the large charge behavior of the quantum entropy function for the BMPV black hole in two limits, firstly when the angular momentum of the black hole scales uniformly with the rest of the charges, and secondly where the black hole is non-rotating. In particular we constructed new saddle points, enumerated in Table \ref{tab:quantization}, of the quantum entropy function which arise from taking orbifolds of the near horizon geometry of the black hole, and computed their leading and next-to-leading contribution to the path integral \eqref{qef}. The contributions to the path integral from these saddle points are exponentially suppressed with respect to the contribution from the leading saddle point, the near horizon geometry itself.
On the microscopic side, an analysis of the appropriate index for BMPV black holes for string theory compactified on a five-torus yields a series of exponentially suppressed terms.
Let us now compare results on the microscopic and macroscopic sides.
\subsection{The Rotating Case}
On the macroscopic side, we have Type (0,0), Type (0,1), Type (1,0) and Type (1,1) orbifolds as in Table \ref{tab:quantization}, and the contribution $\mathcal{Z}|_s$ to the path integral \eqref{qef} from each such orbifold for the case $n_V=27$ is given by\footnote{We used the fact that $A_H\sim \Lambda^{3/2}$.}
\begin{equation}
    \ln\mathcal{Z}|_s = {A_H\over 4s}-6\ln\left( A_H\right)^{2/3}.
\end{equation}
Meanwhile, on the microscopic side we have a class of exponentially suppressed terms with with $\s=1$ and $c>1$. These terms do not carry any arithmetic constraint and appear for all values of charges. Their leading behaviour takes the form\footnote{\label{7} As mentioned above, the next-to-leading term is computed with the assumption that cancellations do not occur between contributions to $\tilde{d}_{micro|s}$.}
\begin{equation}
    \ln\tilde{d}_{micro|c} = {A_H\over 4c}-6\ln\left( A_H\right)^{2/3}.
\end{equation}
Hence the Type (0,0) orbifolds in the bulk are their natural counterpart with $s = \bar s c  = c$, as they appear without arithmetic conditions and reproduce the leading and next-to-leading behaviour correctly. 

We next discuss the contribution for the degeneracy which appear for a given integer $\s >1$, provided the charges satisfy the following quantization conditions: 
\begin{equation}\label{arithmetic}
\s| n \mbox{ and } \s | Q_1Q_5 \mbox{ and }  \left( \s| J\pm 2 \mbox{ or } \s | J\pm 1\mbox{ or } \s | J  \right) ,
\end{equation} 
and their leading behaviour is of the form
\begin{equation}
    \ln\tilde{d}_{micro|{\bar{s}}} = {A_H\over 4\bar{s}}-6\ln\left( A_H\right)^{2/3}.
\end{equation}
We now discuss the macroscopic counterparts of these terms. Firstly we note that to the extent indicated by our analysis, the flux quantization conditions on the orbifolds do not seem to restrict $J$. However, since the quantum entropy function picks out the microcanonical ensemble, we can directly fix the $J$ charge to obey the above condition. With this input, we see that the Type (1,0) and Type (1,1) orbifolds exist only when the above arithmetic conditions are satisfied, with $Q_5$ divisible by $\bar{s}$. Further, they have the same leading and next-to-leading behavior as the microscopic term at hand. This makes them natural counterparts for this term. 

While there are no obvious counterparts of Type (0,1) in the microscopic formulae, this is not necessarily a mismatch. For example, it might as well be that the $\bar s =1$ sector in the microscopic formulae contains the contributions of these orbifolds. A similar statement may hold for the Type (1,0), Type (1,1) orbifolds if $J$ is does not satisfy \eqref{arithmetic}.

\subsection{The Non-Rotating Case} 
On the macroscopic side we have the Type (0,0) orbifolds with the large charge behaviour
\begin{equation}
    \ln\mathcal{Z}|_s = {A_H\over 4s}-\frac{15}{2}\ln\left( A_H\right)^{2/3}.
\end{equation}
Again, one is naturally led to identify the contribution of these saddles to that of the microscopic terms with $\bar{s}=1$ and $c>1$ and hence $s = \bar s c =  c$. However, their leading behaviour is of the form
\begin{equation}
    \ln\tilde{d}_{micro|{\bar{s}}} = {A_H\over 4s}-6\ln\left( A_H\right)^{2/3}.
\end{equation}
It would be interesting to reconcile this mismatch. 

We next consider the Type (0,1), (1,0) and (1,1) orbifolds with the behaviour
\begin{equation}
    \ln\mathcal{Z}|_s = {A_H\over 4s}-6\ln\left( A_H\right)^{2/3},
\end{equation}
and note that on the microscopic side we have terms which appear when the charges satisfy the arithmetic properties
\begin{equation}\label{arithmeticnonrot}
\s| n \mbox{ and } \s | \left(Q_1Q_5\right) ,
\end{equation}
and have the leading behaviour
\begin{equation}
     \ln\tilde{d}_{micro|{\bar{s}}} = {A_H\over 4\bar{s}}-6\ln\left( A_H\right)^{2/3}.
\end{equation}
Hence these terms are the natural counterparts of the above orbifolds, following the discussion in the rotating case.
\subsection{Conclusions}
On comparing the microscopic and macroscopic sides, we see that the situation is only partially satisfactory. In the rotating case, we match the leading and next-to-leading behaviours across an infinite set of saddle points when the black hole charges obey no special quantization condition. This is a definite success. 

Nonetheless, there are some puzzles which appear when we study saddle points which correspond to special arithmetic properties of charges. Firstly, the quantization condition \eqref{arithmetic} on $J$ which appears on the microscopic side does not appear on the microscopic side. It would be interesting to examine the orbifold geometries we have constructed to see if these quantization conditions may indeed be realized in some way. 

Next, in the non-rotating case it is puzzling that though the leading term matches on both sides when there are no arithmetic conditions, the next to leading term does not. In particular, the degeneracy for the horizon is slightly smaller than what is computed from the microscopic counting. When the charges obey arithmetic properties, then the leading and next-to-leading terms match on the microscopic and macroscopic sides, which is satisfactory.

In both the rotating and non-rotating cases, the quantization condition which appears on the D-brane charges $Q_1$ and $Q_5$ in \eqref{arithmetic} and \eqref{arithmeticnonrot} only appears as a condition on $Q_5$ in the macroscopic side. To this order in the large-charge expansion, there does not seem to be a conflict, as it may be that the contribution of microscopic terms when $\bar{s}|Q_1$ and not $Q_5$ is simply contained in the Type (0,0) orbifolds. One would need to compute to higher orders to settle this question.

Finally, another reason for the fact that the match we observe is only partial could also be that the microscopic index contains contributions from $\tfrac18$-BPS states which do not contribute to the microscopic degeneracy of the single-centre black hole, which is what the quantum entropy function measures. It would be of interest to examine if the match becomes exact once this possibility is accounted for on the microscopic side.
\section*{Acknowledgements}
We would like to thank Jin-Beom Bae, Davide Cassani, Atish Dabholkar, Justin David, Euihun Joung, Rajesh Gopakumar, Nick Halmagyi, Sameer Murthy, and especially Ashoke Sen for valuable discussions and comments. SL's work is supported by a Marie Sklodowska Curie Individual Fellowship 2014. PN acknowledges ICTS-TIFR Bengaluru and also acknowledges his debt to the people of India for their steady and generous support to research in the basic sciences. 
\appendix
\section*{Appendix}
\section{Harmonics on AdS$_2$}\label{ads2app}
In this Appendix we will review some essential facts about eigenfunctions of the Laplacian on AdS$_2$ for vector, spin $\frac32$ and spin 2 fields. This has been extensively studied in \cite{Camporesi:1990wm,Camporesi:1994ga,Camporesi:1995fb,Camporesi} and reviewed in \cite{Banerjee:2011jp} and we refer the reader to those papers for more detailed discussions. Further, since we only encounter discrete modes of the Laplacian in our analysis, we will be entirely focused on those modes here. We work in AdS$_2$ coordinates in which the metric is given by
\begin{equation}
ds^2 = a^2\left(d\eta^2 +\sinh^2\eta\,d\theta^2\right),\, \eta\in\left[0,\infty\right),\,\theta\in\left[0,2\pi\right).
\end{equation}
Then the set of discrete modes of the vector field on AdS$_2$ are given by
\begin{equation}\label{vectordiscrete}
f_a^m=\nabla_a\phi^m,\quad \phi^m={\sqrt{1\over 2\pi\vert m\vert}}\left[\sinh\eta\over 1+\cosh\eta\right]^{\vert m\vert}e^{i m\theta},\, m\in \mathbb{Z}-\lbrace0\rbrace.
\end{equation}
Note that though $\phi^m$ is not a normalizable mode on AdS$_2$ and is therefore not included in the scalar heat kernel, $f_a^m$ is a Kronecker delta function normalizable mode of the vector field and hence should be included in the vector heat kernel \cite{Camporesi:1994ga}. Similarly, there exists a discrete set of eigenmodes of the spin-2 Laplacian on $\text{AdS}_2$, with $-\square$ eigenvalue $2\over a^2$. These are modes given by
\begin{equation}\label{metricdiscrete}
\begin{split}
w_{\ell,mn}dx^mdx^n ={a\over\sqrt{\pi}}&\sqrt{\vert\ell\vert\l(\ell^2-1\r)\over 2} {\l(\sinh\eta\r)^{\vert\ell\vert-2}\over\l(1+\cosh\eta\r)^{\vert\ell\vert}}e^{i\ell\theta}\times\\&\times \l(d\eta^2+2i\sinh\eta d\eta d\theta -\sinh^2\eta d\theta^2\r),\quad \ell\in\mathbb{Z},\,\vert\ell\vert\geq 2.
\end{split}
\end{equation}
Finally, we enumerate the discrete modes of the gravitino. These are Kronecker delta function normalizable modes denoted by $\xi$ and $\hat{\xi}$ on AdS$_2$ whose explicit forms are given below. If the two dimensional Dirac matrices are chosen to be
\begin{equation}
\gamma_0=-\tau_2,\quad \gamma_1=\tau_1,
\end{equation}
where the $\tau_i$s are the Pauli matrices, the $\xi$ spinors are given by
\begin{equation}
\xi_{\eta}^{(k)\pm} = {1\over 8\pi a}\,
e^{ i\left(k+{1\over 2}\right)\phi} 
\left(\begin{array}{cc} 0 \\ \pm {\sinh^{k-1}{\eta\over 2}\over\cosh^{k+2} {\eta\over 2}} \end{array}\right),\quad \xi_{\theta}^{(k)\pm} = {1\over 4\pi a}\,
e^{ i\left(k+{1\over 2}\right)\phi} 
\left(\begin{array}{cc} 0 \\ \pm {\sinh^{k}{\eta\over 2}\over\cosh^{k+1} {\eta\over 2}} \end{array}\right),
\end{equation}
and the hatted spinors are given by
\begin{equation}
\hat{\xi}_{\eta}^{(k)\pm} = {1\over 8\pi a}\,
e^{ -i\left(k+{1\over 2}\right)\phi} 
\left(\begin{array}{cc} \pm {\sinh^{k-1}{\eta\over 2}\over\cosh^{k+2} {\eta\over 2}}\\0 \end{array}\right),\quad \hat{\xi}_{\theta}^{(k)\pm} = {1\over 4\pi a}\,
e^{ -i\left(k+{1\over 2}\right)\phi} 
\left(\begin{array}{cc} \pm {\sinh^{k}{\eta\over 2}\over\cosh^{k+1} {\eta\over 2}}\\0 \end{array}\right).
\end{equation}
\section{Flux Quantization for Orbifolds}\label{Flux_Quantization}
Since the flux is a 3-form, we shall consider the integral
\begin{equation}
 \int_{\text{3 Cycle}} F
\end{equation}
over all non-contractible 3-cycles in the geometry which are left invariant by the action of the $\mathbb{Z}_s$ orbifold. It is straightfoward to firstly observe that the 3-cycle must be located at $\eta=0$. At this point, the flux $F_3$ takes the form
\begin{equation}
   F_{3}= - \frac{Q_5}{4} \sin \psi  \ d\phi \wedge d\psi  \wedge dx_4 -\frac{Q_5 J  \sin \psi \  d\phi \wedge d\psi \wedge d\chi }{32 r_0^2}+\ldots,
\end{equation}
where the `$\ldots$' denote components of the flux which are along the $\eta$ or $\theta$ directions. Of these, the first term is proportional to the volume form of $S^3$ and gives rise to the flux quantization condition
\begin{equation}
    Q_5\in \mathbb{Z}.
\end{equation}
This is because all the $\mathbb{Z}_s$ orbifolds we construct reduce this $S^3$ to $\tfrac{1}{s}$ of its original size.
The second term doesn't contribute a quantization condition as the 2-cycle $d\phi \wedge d\psi$
is contractible in $S^3$.

\bibliographystyle{JHEP}
\bibliography{paper}
\end{document}